\documentclass[conference]{IEEEtran} 
\usepackage{graphicx}
\usepackage{amsmath}
\usepackage{amssymb}
\usepackage{mathrsfs}
\usepackage{stfloats}
\usepackage{dsfont}
\usepackage{setspace}
\usepackage{epstopdf}
\usepackage{cite}
\usepackage{balance}
\usepackage{xcolor}
\usepackage{booktabs}
\usepackage{pifont}
\usepackage{algorithm}

\usepackage{accents}

\makeatletter
\newcommand{\doublewidetilde}[1]{{%
		\mathpalette\double@widetilde{#1}%
}}
\newcommand{\double@widetilde}[2]{%
	\sbox\z@{$\m@th#1\widetilde{#2}$}%
	\ht\z@=.5\ht\z@
	\widetilde{\box\z@}%
}
\makeatother

\begin{document}

\title{
Secrecy Performance of RIS-assisted Wireless Networks under Rician fading
}

\author{

\IEEEauthorblockN{Thi Tuyet Hai Nguyen and Tien Hoa Nguyen*}
\IEEEauthorblockA{\textit{Faculty of Information Technology 2, Posts and Telecommunications Institute of Technology, Ho Chi Minh, Viet Nam}\\
\textit{School of Electronics and Telecommunications, Hanoi University of Science and Technology, Hanoi 100000, Vietnam} \\
tuyethai@ptithcm.edu.vn; hoa.nguyentien@hust.edu.vn}


%
}
\maketitle
%
\begin{abstract}
	Secrecy outage probability (SOP) and secrecy rate (SR) of the reconfigurable intelligent surface (RIS) assisted wireless networks under Rician fading are investigated in this paper. More precisely, we enhance the secrecy performance of the considered networks by suppressing the wiretap channel instead of maximizing the main channel. We propose a simple heuristic algorithm to find out the optimal phase-shift of each RIS's element. Simulation results based on the Monte-Carlo method are given to verify the superiority of the proposed optimal phase-shifts compared to the random phase-shifts design.
\end{abstract}
\section{Introduction}
It is expected that the number of wireless devices connecting to the Internet will be up to 28.5 billion by 2022 according to Cisco's report \cite{Report:01}.
Such ultra-dense devices networks, however, pose many challenges to the current infrastructure, i.e., how to supply energy to feed such networks, how to improve the spectral efficiency of the whole networks and how to better protect networks from cyber-attack and so on.
Considering the security perspective, due to the broadcast nature of the wireless medium, wireless communications are susceptible to cybercriminal activities. The conventional approach to deal with the cyber-attack is to employ high complexity cryptography technology.
Nonetheless, it generally requires high-cost hardware in order to support these expensive computations which are not suitable for the low-cost devices such as low power wide area networks (LPWAN) devices \cite{Raza:LoRa:01,Thanh:LoRa:01}.
Fortunately, physical layer security (PLS) has recently emerged as a cost-effective solution to ameliorate network security from the information-theoretic aspect.
More precisely, PLS exploits the physical characteristics of the wireless channel for securely transmitting messages.
Reconfigurable intelligent surface (RIS), on the other hand, is a novel technology and is considered one of the most promising technologies to enhance both energy and spectral efficiency. 
The main advantage of RIS lies in its ability to reconfigure the incident signal through passive beamforming, which controls the phase of the signal. 
This, as a result, permits an effective way to control the radio waves
for extending coverage area and increasing the data rate \cite{Basar:RIS:01}.

The performance of RIS and PLS itself was studied widely in the literature \cite{Duy:PHY:02, Duy:PHY:01, Phong:RHY:01, Chien:RIS:01, Chien:RIS:02}.
More precisely, secrecy outage probability (SOP) of the dual-hop cooperative networks with the presence of the co-channel interferer was derived in \cite{Duy:PHY:02}. 
Duy \textit{et al.} in \cite{Duy:PHY:01} studied the SOP and secrecy capacity (SC) by employing the $k$-th best relay selection. 
In \cite{Phong:RHY:01}, the SOP of the cognitive radio networks (CRNs) under Nakagami-$m$ fading was computed in closed-form expression.
Coverage probability (Pcov) and average rate of the RIS-aided wireless networks under both correlated and uncorrelated Rayleigh fading were derived in \cite{Chien:RIS:01, Chien:RIS:02}.
Particularly, the closed-form expressions of these metrics were obtained and the impact of the number of RIS's elements on the performance of the Pcov was identified as well.
The performance of the combination of RIS and PLS was also given in \cite{Yang:RIS-PHY:01, Trigui:RIS-PHY:01, Khoshafa:RIS-PHY:01}.
To be more precise, Yang \textit{et al.} studied the SOP by approximating the received signal of the indirect link as a Gaussian random variable.
The work in \cite{Trigui:RIS-PHY:01}, on the other hand, derived the SOP based on a large number of RIS elements assumption.
Authors in \cite{Khoshafa:RIS-PHY:01} derived the SOP, probability of non-zero of cellular networks and outage probability of device-to-device (D2D) communications.  

Different from the above-mentioned works, in the present paper we enhance the secrecy performance of the RIS-assisted wireless networks by minimizing the wiretap links under the Rician fading. More precisely, the principal novelties and contributions are given as follows:
\begin{itemize}
    \item The performance of both SOP and secrecy rate (SR) is investigated under the Rician fading.
    \item We formulate a minimization problem of the end-to-end (e2e) signal-to-noise ratios (SNRs) of the eavesdropper channel. 
    A simple heuristic algorithm is provided to solve the considered problem.
    \item Monte-Carlo simulations are given to verify the superior of the suppression of the eavesdropper channel compared to the random phase-shift setup.
    \item Numerical results showed that increasing transmit power, number of the RIS's elements are beneficial for improving system performance.
\end{itemize}
\section{System model}
Let us consider a wireless networks comprising a source node (denoted by S), a legitimate destination (denoted by D), an active eavesdropper (denoted by E) and a RIS with $\mathcal{K}$ elements. The source node transmits confidential information to the main destination via the help of the RIS. The eavesdropper, on the other hand, attempts to eavesdrop this information.
We assume that the source node has the perfect channel state information (CSI) of all transmission links (including the wiretap link owing to the active eavesdropper).
Additionally, we also assume that the direct link from S to D and E is blocked due to the long transmission distance \cite{Thanh:COOP:01, Hoa2020}.
\subsection{Channel modelling}
In this work, all transmission links are suffered from both large-scale path-loss and small-scale fading. 
The impact of the shadowing does not take into consideration since it is a general case in the literature \cite{Marco:SG:01, Hoa2021}.
\subsubsection{Small-scale fading}
Let us denote $\mathbf{g}_{u,v} \in \mathbb{C}^\mathcal{K}$ as the channel coefficient from node $u \in \left\{ S, R \right\}$
to node $v \in \left\{ D, E, R \right\}$ and follows by a complex Gaussian distribution with $\mu_{u,v}$ mean and $\eta_{u,v}$ variance, i.e., ${\mathbf{g}_{u,v}} \sim \mathcal{CN}\left( \mathbf{\mu}_{u,v}, \mathbf{\eta}_{u,v} \mathbf{I} \right)$. $\mathbf{I}$ is the unitary matrix with size $\mathcal{K}$. Here $\mu_{u,v}$ and $\eta_{u,v}$ are a function of both the ratio of the line-of-sight (LOS) to the non-LOS (NLOS) component denoted by $K_{u,v}$ and the large-scale path-loss which is given in the sequel.
\subsubsection{Large scale path-loss}
Let us denote $L_{u,v}$ as the large-scale path-loss from $u$ to $v$ and is formulated as \cite{Thanh:SG:01}
\begin{align}
    L_{u,v} = K_0 \left( 1 + {d_{u,v}} \right)^\beta,
\end{align}
where $\beta$ and $K_{0} = \left( 4\pi {f_c} / c \right)^{2}$ are the path-loss exponent and the path-loss constant, respectively;
$c = 3 \times 10^8$ (in meters per second) is the speed of light and $f_c$ is the carrier frequency (in Hz). 
It is noted that in this paper we consider the bounded path-loss model in place of the unbounded path-loss model. The main advantage of the considered model is that it overcomes the singularity issue of the unbounded model when the receiver is in proximity to the transmitter \cite{Jian:UAV:01}.

The received signal at node D and E denoted by $y_D$ and $y_E$ is then given as
\begin{align} \label{Eq:Y_D:01}
    y_D = \sqrt{P_{\rm{tx}} } \mathbf{g}_{S,R}^H \pmb{\Theta} \mathbf{g}_{R,D} + n_D,
    \\ 
    \label{Eq:Y_D:01}
    y_E = \sqrt{P_{\rm{tx}} } \mathbf{g}_{S,R}^H \pmb{\Theta} \mathbf{g}_{R,E} + n_E,
\end{align}
where $P_{\rm{tx}}$ is the transmit power of $S$,
$n_D$ and $n_E$ are the additive white Gaussian noise (AWGN) of D and E. 
$\pmb{\Theta} \in \mathbb{C}^{\mathcal{K} \times \mathcal{K}}$ is the phase-shift matrix and is given as $\pmb{\Theta} = \mathrm{diag}\big([e^{j\omega_{1}}, \ldots, e^{j\omega_{ \mathcal{K} }}]^T \big)$, where $\omega_{k} \in [-\pi, \pi], k \in \left\{ 1, \ldots, \mathcal{K} \right\}$ is the phase-shift of the $k-$th reflecting element;
$\mathrm{diag} \left( . \right)$ is the diagonal matrix;
$\left( . \right)^H$ is the conjugate transpose; $\left( . \right)^T$ is the transpose;
and $j$ is the imaginary unit.
\subsection{Signal-to-noise ratios (SNRs)}
The signal-to-noise ratios (SNRs) of node $o \in \left\{ D, E \right\}$ denoted by ${\rm{SNR}}_o$ is then formulated as
\begin{align}
    {\rm{SNR}}_o =& 
    \frac{ P_{\rm{tx}}  \left|\mathbf{g}_{S,R}^H \pmb{\Theta} \mathbf{g}_{R,o} \right|^2 }{ \sigma_o^2 }.
\end{align}
Here $\sigma_o^2$ is the noise variance of node $o$ and is computed as $\sigma_o^2 = \sigma^2 = -174 + \text{NF} + 10 \log \left( \text{Bw} \right), \forall o$ (in dBm) \cite{Thanh:LoRa:02}, Bw (in Hz) is the transmission bandwidth and NF (in dB) is the noise figure.
\section{Performance Metrics}
We study the performance of two vital metrics of PLS, i.e., the SOP and the SR.
The SOP measures the probability that the secrecy rate is lower than a predefined threshold.
The SR is the maximal transmission rate that the eavesdropper is not able to successfully decode any information.
Mathematical speaking, they are given as
\begin{align} \label{Eq:SOP_SC:01}
    \rm{SOP} =& \Pr \left\{ \log_2 \left( \frac{1 + {\rm{SNR}}_D }{1 + {\rm{SNR}}_E  } \right) \le R \right\}
     \\
    \rm{SR} =& \mathbb{E} \left\{ \max \left\{ {{{\log }_2}\left( 1 + {\rm{SNR}}_D \right) - {{\log }_2} \left( 1 + {\rm{SNR}}_E \right),0} \right\} \right\},
    \nonumber
\end{align}
where $\log$ and $\max$ are the logarithm and maximum function; $R$ (in [bits/s/Hz]) is the expected rate; $\Pr \left\{ . \right\}, \mathbb{E} \left\{ . \right\}$ are the probability and expectation operators.
\subsection{Optimal phase-shifts design}
Different from other works in the literature we are going to suppress the wiretap channel in place of maximizing the main channel \cite{Khoshafa:RIS-PHY:01, Trigui:RIS-PHY:01, Yang:RIS-PHY:01}.
More precisely, we formulate the following minimization problem 
\begin{align} \label{Eq:prob:01}
    \mathop {\min }\limits_{{\omega _k} \in \left[ { - \pi ,\pi } \right],\forall k}  |\mathbf{g}_{S,R}^H \pmb{\Theta} \mathbf{g}_{R,E}|^2.
\end{align}
It is no doubt that the minimization problem in \eqref{Eq:prob:01} is non-convex due to the nonlinearity of the utility function. We thus propose a simple heuristic algorithm to find the optimal phase shifts of each RIS element.
Particularly, we first start computing the absolute of $\mathbf{g}_{S,R}, \mathbf{g}_{R,E}$. 
We then sort the product of the absolute of $\mathbf{g}_{S,R}, \mathbf{g}_{R,E}$ in the descending order, e.g., $| {g_{S,R}^{\left( k \right)}} | | {g_{R,E}^{\left( k \right)}} |$, where $\left( 1 \right) > \cdots > \left( k \right) > \cdots > \left( \mathcal{K} \right)$.
Next, we split the sorted vector in the previous step into $\mathcal{N} = \left\lfloor { \mathcal{K}/2} \right\rfloor$ pairs denoted by ${s_n} = \left\{ {| {g_{S,R}^{\left( 2n-1 \right)}} || {g_{R,E}^{\left( 2n-1 \right)}} |,| {g_{S,R}^{\left( 2n \right)}} || {g_{R,E}^{\left( 2n \right)}} |} \right\}$, $n \in \left\{ 1, \ldots, \mathcal{N} \right\}$. $\left\lfloor . \right\rfloor $ is the floor function.
Finally, the optimal phase-shift of $\left( n \right)$-th element denoted by $\omega_{\left( n \right)}^*$ is the root of following system equations
\begin{align} \label{Eq:Opt:01}
    & \arg \left( g_{S,R}^{\left( {2n - 1} \right)} \right) +
    \arg \left( g_{R,E}^{\left( {2n - 1} \right)} \right) +   \arg \left( { \omega_{\left( {2n - 1} \right)}^* } \right) 
    \nonumber \\
    &= \pi  + \arg \left( g_{S,R}^{\left( {2n} \right)} \right) + \arg \left(  g_{R,E}^{\left( {2n} \right)} \right) + \arg \left( {\omega_{\left( {2n} \right)}^* } \right), \forall n,
\end{align}
where $\arg \left( x \right)$ is the angle of the complex number $x$.
The summary of the above steps is provided in Algorithm \ref{Algorithm1}. Having obtained the optimal phase-shifts of RIS, we plug it into \eqref{Eq:SOP_SC:01} for computing the SOP and SC.
%
\begin{algorithm}[!ht] 
	\caption{A simple heuristic algorithm to find the optimal phase-shifts of RIS} \label{Algorithm1}
	\textbf{Input}: $\mathbf{g}_{S,R}, \mathbf{g}_{R,E}$.
	\begin{itemize}
		\item[1.] Compute the modulus of each element of $\mathbf{g}_{S,R}, \mathbf{g}_{R,E}$ denoted by $| g_{\{S,R\},k}|$,$| g_{\{R,E\},k}|$.
		\item[2.] Sort the product of the absolute of $\mathbf{g}_{S,R}, \mathbf{g}_{R,E}$ in downward order, i.e., $| {g_{S,R}^{\left( k \right)}} | | {g_{R,E}^{\left( k \right)}} |$, 
		$| {g_{S,R}^{\left( 1 \right)}} || {g_{R,E}^{\left( 1 \right)}} | > \cdots > | {g_{S,R}^{\left( k \right)}} || {g_{R,E}^{\left( k \right)}} | > \cdots > | {g_{S,R}^{\left( \mathcal{K} \right)}} | | {g_{R,E}^{\left( \mathcal{K} \right)}} |$.
		\item[3.] Divide the sorted vector in step 2 into $\mathcal{N} = \left\lfloor {\mathcal{K}/2} \right\rfloor $ pairs and denoted by $S = \left\{ {{s_1},...,{s_\mathcal{N}}} \right\}$ where ${s_n} = \left\{ {| {g_{S,R}^{\left( 2n-1 \right)}} || {g_{R,E}^{\left( 2n-1 \right)}} |,| {g_{S,R}^{\left( 2n \right)}} || {g_{R,E}^{\left( 2n \right)}} |} \right\}$, $n \in \left\{ 1, \ldots, \mathcal{N} \right\}$.
		\item[4.] Solve the system equations in \eqref{Eq:Opt:01} to obtain the optimal phase shift $\omega_{\left( n \right)}^*$, $n \in \left\{ 1, \ldots, \mathcal{N} \right\}$.
	\end{itemize}
	\textbf{Output}: The optimal phase of all RIS's elements.
\end{algorithm}
\subsection{Random phase-shift design}
Under the random phase-shift scheme, the phase of each RIS element is randomly assigned from $-\pi$ to $\pi$, i.e., $\omega_k \in \left[ -\pi,\pi \right], \forall k$, it then substitutes into \eqref{Eq:SOP_SC:01} to compute the SOP and SR.
%
\section{Numerical Results}
In this section, simulation results based on the Monte-Carlo method are given to investigate the performance of the considered systems. 
Unless otherwise stated, following parameters are employed throughout this section: $\mathcal{K} = 144$; $R = 1$ bits/s/Hz; $P_{\text{tx}} = 20$ dBm; NF = 6 dB; Bw = 10 MHz; $\beta = -2.5$; $f_c = 2.1$ GHz; 
${\mu _{u,v}} = \sqrt {{K_{uv}}{L_{u,v}}} ;{\eta _{u,v}} = {L_{u,v}}$; $K_{\rm{sr}} = 3$, $K_{\rm{rd}} = 0.5$; $K_{\rm{re}} = 1.25$;
the position of $S, R, D$ and $E$ are $\left( 0,0 \right)$, $\left( 10, 10 \right)$, $\left( 70, 0 \right)$ and $\left( 70, -10 \right)$, respectively.
\begin{figure}[!ht]
    \centering
    \includegraphics[width=0.4\textwidth]{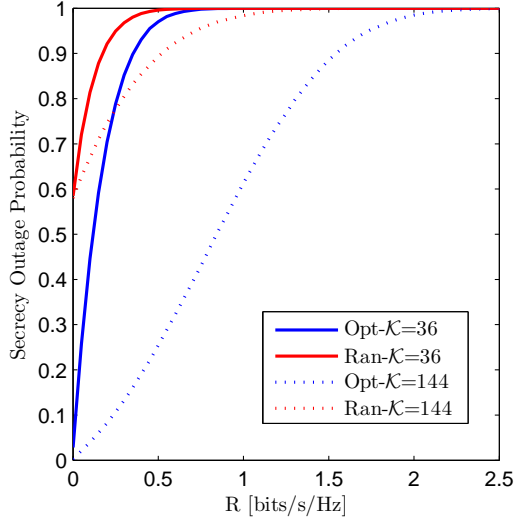}
    \caption{
           SOP vs. $R$ [bits/s/Hz] with various value of $\mathcal{K}$.
				} 
    \label{fig:SOP_Rate:01}
\end{figure}

Fig. \ref{fig:SOP_Rate:01} unveils that the SOP of both schemes, namely, optimal and random phase shifts denoted by ``Opt" and ``Ran" goes up with the increase of $R$. The explanation is directly obtained from the definition of the SOP.
Moreover, it is apparent that the ``Opt" scheme outperforms its counterpart especially when $\mathcal{K} \gg 1$.
Particularly, the `Opt" curve is almost $2\times$ better than the ``Ran" scheme with $\mathcal{K} = 144$ and $R = 1$ bits/s/Hz.
Additionally, increasing $\mathcal{K}$ is beneficial for both schemes.
Interestingly, under the ``Ran" scheme, the SOP does not begin from zero even with $R = 0$. It means that the perfect secrecy transmission does not exist under this scheme regardless of the value of $R$ and this is contrary to the proposed solution.

\begin{figure}[!ht]
    \centering
    \includegraphics[width=0.4\textwidth]{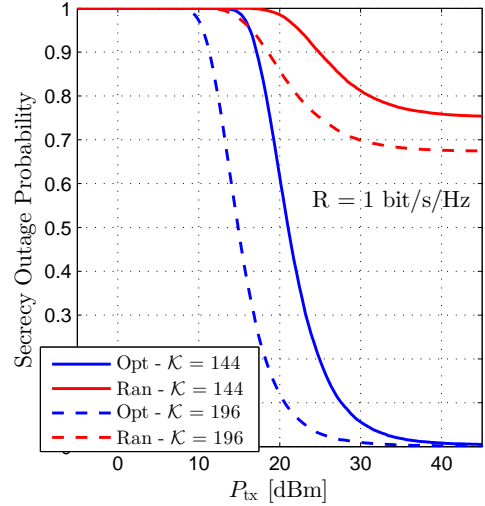}
    \caption{
           SOP vs. $P_{\text{tx}}$ [dBm] with different RIS elements $\mathcal{K}$.
				} 
    \label{fig:SOP_Ptx:01}
\end{figure}

\begin{figure}[!ht]
    \centering
    \includegraphics[width=0.4\textwidth]{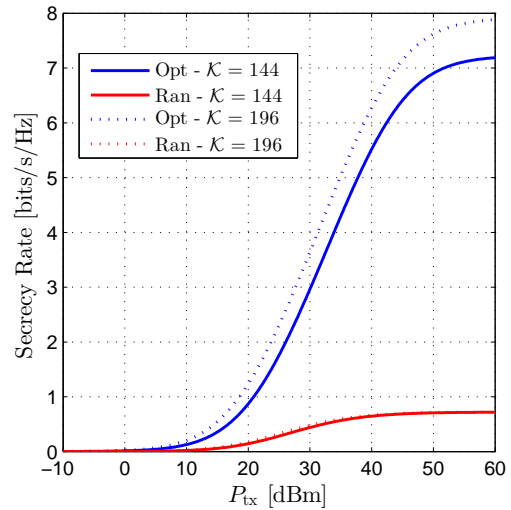}
    \caption{
           SR vs. $P_{\text{tx}}$ [dBm] with different RIS elements $\mathcal{K}$.
				} 
    \label{fig:SC_Ptx:01}
\end{figure}

Figs. \ref{fig:SOP_Ptx:01} and \ref{fig:SC_Ptx:01} stretch the performance of SOP and SR versus the transmit power. We see that the SOP and the SR experience different behaviors with respect to the $P_{\rm{tx}}$. Particularly, SOP is a monotonic decreasing function while SR is a monotonic increasing. 
However, there exist a bound for both metrics, e.g., a lower bound for the SOP and an upper bound for the SR. It can be explained that under the high transmit power regime we can employ following approximation ${\log _2}\left( {\frac{{1 + \frac{{{P_{{\rm{tx}}}}{{\left| {{\bf{g}}_{S,R}^H \rm{\pmb{\Theta}} {{\bf{g}}_{R,D}}} \right|}^2}}}{{  \sigma _D^2}}}}{{1 + \frac{{{P_{{\rm{tx}}}}{{\left| {{\bf{g}}_{S,R}^H\rm{\pmb{\Theta}} {{\bf{g}}_{R,E}}} \right|}^2}}}{{ \sigma _E^2}}}}} \right)\mathop  = \limits^{{P_{{\rm{tx}}}} \gg 1} {\log _2}\left( {\frac{{{{\left| {{\bf{g}}_{S,R}^H\rm{\pmb{\Theta}} {{\bf{g}}_{R,D}}} \right|}^2}}}{{{{\left| {{\bf{g}}_{S,R}^H\rm{\pmb{\Theta}} {{\bf{g}}_{R,E}}} \right|}^2}}}} \right)$ which is independent of $P_{\rm{tx}}$ thus increasing $P_{\rm{tx}}$ is not always beneficial. Once again, we observe that the proposed optimal phase-shift designs is far better than the random scheme. For example, the SR of the optimal scheme achieves almost 8 bits/s/Hz while the SR of the random one attains just below 1 bit/s/Hz for $\mathcal{K} = 196$. 

\begin{figure}[!ht]
    \centering
    \includegraphics[width=0.4\textwidth]{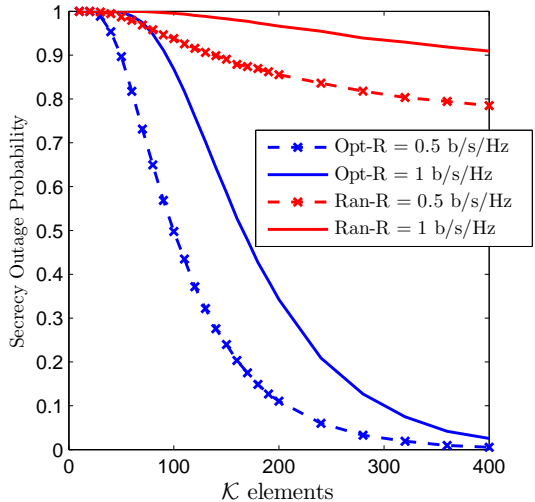}
    \caption{
           SOP vs. $\mathcal{K}$ elements with various value of $R$.
				} 
    \label{fig:SOP_N:01}
\end{figure}

\begin{figure}[!ht]
    \centering
    \includegraphics[width=0.4\textwidth]{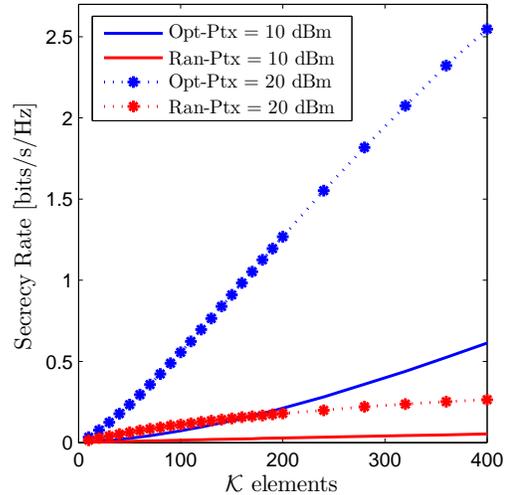}
    \caption{
           SR vs. $\mathcal{K}$ elements with various value of the transmit power $P_{\rm{tx}}$.
				} 
    \label{fig:SC_N:01}
\end{figure}

Figs. \ref{fig:SOP_N:01} and \ref{fig:SC_N:01} illustrate the impact of the number of RIS elements $\mathcal{K}$ on the performance of the SOP and SR. 
We observe that Fig. \ref{fig:SOP_N:01} confirms the comments in Fig. \ref{fig:SOP_Rate:01} that scaling up $\mathcal{K}$ improves the system performance.
More precisely, SOP continuously declines from 1 to 0 for the ``Opt" curve with $R$ = 0.5 bits/s/Hz when $\mathcal{K}$ goes from 10 to 400.
The ``Ran" scheme, however, only reduces to 0.8 when $\mathcal{K}$ reaches 400.
It, as a consequence, highlights the benefits of minimizing the eavesdropper link along with the maximization of the legitimate channel.
Besides, Fig. \ref{fig:SC_N:01} shows the secrecy rate regarding to the number of RIS elements $\mathcal{K}$ with various value of the transmit power $P_{\rm{tx}}$. 
It is evident that the optimal scheme always outperforms than another. Additionally, despite of scaling up $\mathcal{K}$ improving system performance the ``Opt" scheme ameliorates much higher than the ``Ran" scheme especially when $P_{\rm{tx}}$ is large.

\begin{figure}[!ht]
    \centering
    \includegraphics[width=0.4\textwidth]{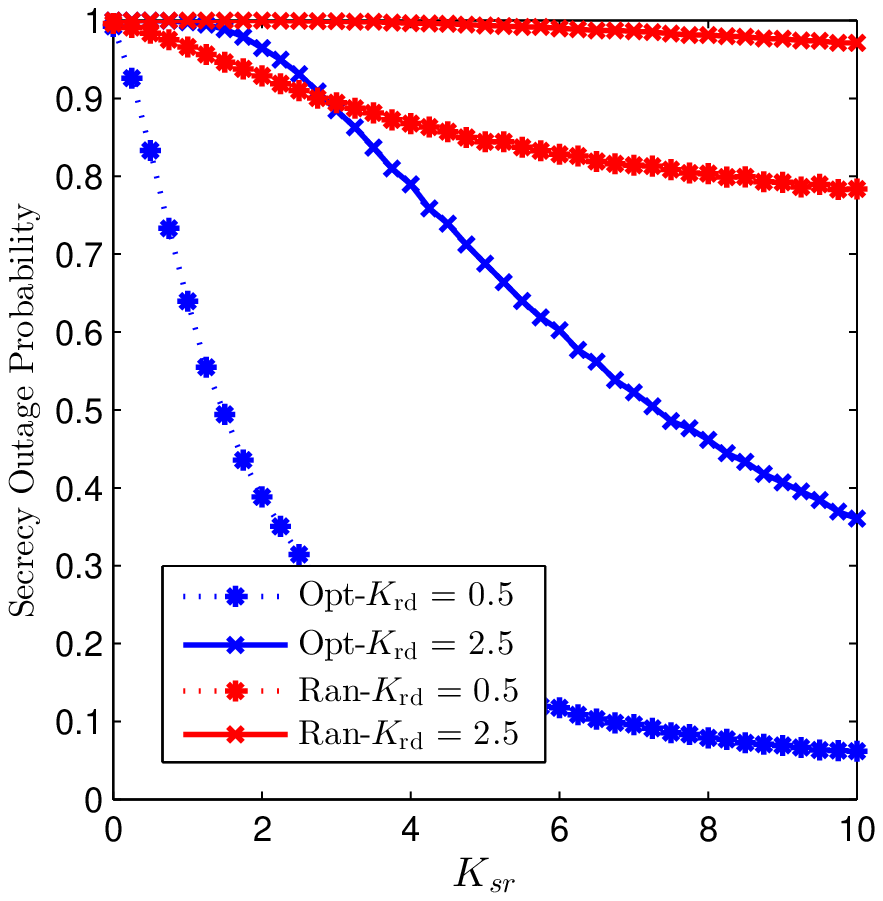}
    \caption{
          SOP vs. $K_{\rm{sr}}$ with various value of $K_{\rm{rd}}$.
				} 
    \label{fig:SOP_xR:01}
\end{figure}
\begin{figure}[!ht]
    \centering
    \includegraphics[width=0.4\textwidth]{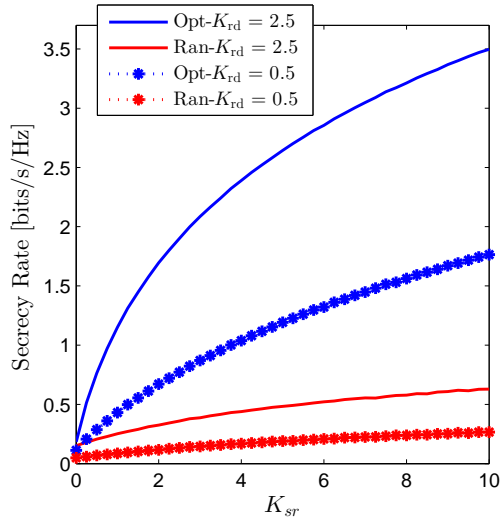}
    \caption{
          SR vs. $K_{\rm{sr}}$ with various value of $K_{\rm{rd}}$.
				} 
    \label{fig:SC_xR:01}
\end{figure}

Figs. \ref{fig:SOP_xR:01}  and \ref{fig:SC_xR:01} address the impact of  $K_{\rm{sr}}$ on the performance of SOP and SR. It is obvious that increasing  $K_{\rm{sr}}$ will improve the performance of both metrics thanks to the increase of the LOS component. However, the improvement of the ``Ran" scheme is a minority compared with the ``Opt".
To be more precise, the SOP of the ``Ran" scheme of  $K_{\rm{rd}}$ is almost stable with the increase of $K_{\rm{sr}}$ while the SOP of the ``Opt" scheme declines significantly when $K_{\rm{sr}}$ goes from 0 to 10. 
\section{Conclusion}
In the present paper, we investigated the secrecy performance of wireless networks with the aid of RIS. We showed that under the active eavesdropper scenario, by minimizing the received signal at the eavesdropper we were able to enhance significantly the system performance. We also provided the results under the random phase-shift scenario that acts as the benchmark of the proposed solution. Our findings showed that increasing the transmit power, the number of RIS elements is beneficial for ameliorating the system performance.
\section*{Acknowledgement}
This research is funded by Hanoi University of Science and Technology (HUST) under project number T2021-SAHEP-002


\end{document}